\documentclass[conference]{IEEEtran}
\IEEEoverridecommandlockouts
\usepackage{cite}
\usepackage{amsmath,amssymb,amsfonts}
\usepackage{algorithmic}
\usepackage{graphicx}
\usepackage{textcomp}
\usepackage{xcolor}
\def\BibTeX{{\rm B\kern-.05em{\sc i\kern-.025em b}\kern-.08em
    T\kern-.1667em\lower.7ex\hbox{E}\kern-.125emX}}
\begin{document}

\title{Respiratory Disease Classification and Biometric Analysis Using Biosignals from Digital Stethoscopes
\thanks{This research is supported by the Academy of Finland 6G Flagship program (Grant 346208) and PROFI5 HiDyn (Grant 326291).}
}

\author{
\IEEEauthorblockN{Constantino Álvarez Casado$^{\star}$, Manuel Lage Cañellas$^{\star}$, Matteo Pedone$^{\star}$, \\ Xiaoting Wu$^{\star}$, Le Nguyen$^{\star}$, Miguel Bordallo López$^{\star}$$^{\dagger}$}
\IEEEauthorblockA{\textit{$^{\star}$Center for Machine Vision and Signal Analysis (CMVS), University of Oulu} \\
\textit{$^{\dagger}$VTT Technical Research Centre of Finland}\\
Oulu, Finland \\
\{constantino.alvarezcasado, miguel.bordallo\}@oulu.fi}

}

\maketitle

\begin{abstract}

Respiratory diseases remain a leading cause of mortality worldwide, highlighting the need for faster and more accurate diagnostic tools. This work presents a novel approach leveraging digital stethoscope technology for automatic respiratory disease classification and biometric analysis. Our approach has the potential to significantly enhance traditional auscultation practices. By leveraging one of the largest publicly available medical database of respiratory sounds, we train machine learning models to classify various respiratory health conditions. Our method differs from conventional methods by using Empirical Mode Decomposition (EMD) and spectral analysis techniques to isolate clinically relevant biosignals embedded within acoustic data captured by digital stethoscopes. This approach focuses on information closely tied to cardiovascular and respiratory patterns within the acoustic data. Spectral analysis and filtering techniques isolate Intrinsic Mode Functions (IMFs) strongly correlated with these physiological phenomena. These biosignals undergo a comprehensive feature extraction process for predictive modeling. These features then serve as input to train several machine learning models for both classification and regression tasks. Our approach achieves high accuracy in both binary classification (89\% balanced accuracy for healthy vs. diseased) and multi-class classification (72\% balanced accuracy for specific diseases like pneumonia and COPD). For the first time, this work introduces regression models capable of estimating age and body mass index (BMI) based solely on acoustic data, as well as a model for sex classification. Our findings underscore the potential of intelligent digital stethoscopes to significantly enhance assistive and remote diagnostic capabilities, contributing to advancements in digital health, telehealth, and remote patient monitoring.

\end{abstract}


\section{Introduction}
\label{sec:intro}

Respiratory conditions, including Asthma, Chronic Obstructive Pulmonary Disease (COPD), and Pneumonia, significantly impact global health, affecting individuals worldwide and challenging healthcare systems. The importance of prompt diagnosis cannot be overstated, as delays may result in critical health deterioration. This situation is exacerbated by a scarcity of healthcare resources, especially skilled clinicians essential for accurate diagnosis \cite{momtazmanesh2023globalRespiratory}. The rise of telehealth highlights the demand for effective diagnostic technologies. Traditional diagnostic methods, such as chest x-rays, often fall short in sensitivity and are not widely available in low- and middle-income countries (LMICs). On the other hand, stethoscope-based auscultation, which analyzes respiratory sounds, offers a non-invasive, cost-efficient, and longstanding approach to lung disease diagnosis. The technical analysis of these sounds provides a pathway to swift, objective, and accurate diagnostic assessments, improving patient care outcomes \cite{kim2022comingStethoscope}.

Our work introduces a novel approach in automated respiratory diagnostics through a multi-step classification framework based on in-depth analysis of stethoscope-derived audio biosignals. Significantly, we pioneer the use of stethoscope-derived acoustic signals for both biometric classification and regression tasks to infer age, sex, and BMI. Leveraging a large, publicly available respiratory sound database from the International Conference on Biomedical Health Informatics (ICBHI) \cite{rocha2018ICBHIDatabase}, our main contributions include:


\begin{itemize}
\setlength\itemsep{0pt}
\setlength\parskip{0pt}
\item Extraction of clinically relevant biosignals from acoustic stethoscope-derived data using EMD and spectral analysis techniques, and the use of physiological-related features computed from these biosignals to train machine learning models.
\item Two binary classification models: one distinguishes between patients with respiratory conditions and healthy individuals of all ages, achieving a balanced accuracy of 89\%, while the other identifies COPD from other respiratory conditions with a balanced accuracy of 94\%.

\item For the first time, to the best of our knowledge, stethoscope audio signals have been used to estimate key biometric indicators such as age, sex, and BMI, marking a novel contribution in the field.
\end{itemize}

Our work enhances the capabilities of telehealth and assistive diagnosis support in healthcare, offering precise remote diagnostics via non-invasive stethoscope technology.

\section{Related work}
Driven by the advancements in digital stethoscopes, automatic respiratory disease diagnosis using acoustic signals has become an increasingly active area of research \cite{kim2022comingStethoscope}. These sophisticated devices offer enhanced sound quality, digital recording capabilities, and the potential for real-time analysis through embedded algorithms. Notably, their connectivity facilitates data transmission and integration with clinical information systems, opening doors for novel diagnostic approaches and potentially transforming respiratory disease management.

The field of audio-based respiratory condition detection has significantly evolved, primarily driven by advancements in machine learning (ML) and deep learning (DL) techniques. The research has largely focused on identifying abnormal pulmonary sounds such as crackles, found in conditions like lung fibrosis, and wheezes, common in obstructive diseases like asthma and COPD \cite{Reichert2008DSPRespiratory}. Initial efforts hinged on digital signal processing, employing time and frequency domain analyses to isolate these sounds \cite{Reichert2008DSPRespiratory,Fattahi2022Filtering,lin2015automaticDSPResp}. Traditional machine learning algorithms furthered this research, utilizing feature extraction in both time and frequency domains, with a particular emphasis on Mel-frequency Cepstral Coefficients (MFCC). Various methods such as Gaussian Mixture Models with MFCC and FFT features \cite{bahoura2009patternGMMs}, Support Vector Machines using MFCCs  \cite{aykanat2017classificationSVM}, ensemble learning methods like Boosting with Discrete Wavelet Transform and MFCCs \cite{kok2019novelDWTMFCC} and Ensemble of Bagged Trees using EMD and statistical features \cite{Sibghatullah2021COPDdetection}, as well as Linear Predictive Cepstral Coefficient-based features with MultiLayer Perceptron (MLP) \cite{mukherjee2021automaticLPCC}, were among the early techniques applied to respiratory sound classification tasks, particularly using the ICBHI17 respiratory sound database \cite{rocha2018ICBHIDatabase}. With increasing computational power, deep learning models, notably Convolutional Neural Networks (CNNs), have come to the fore. Aykanat et al. \cite{aykanat2017classificationSVM} used spectrogram images to train CNNs , while Bardou et al. \cite{BARDOU2018CNNsResp} utilized a ‘Features+CNN’ approach. Recurrent Neural Networks (RNN), such as LSTM and GRU models, have also shown potential, with attention-based variants yielding excellent results in respiratory condition recognition \cite{wall2022deep}.

\section{Proposed Methodology}
\begin{figure*}[ht!]
 \begin{center}
   \includegraphics*[width=0.99\textwidth]{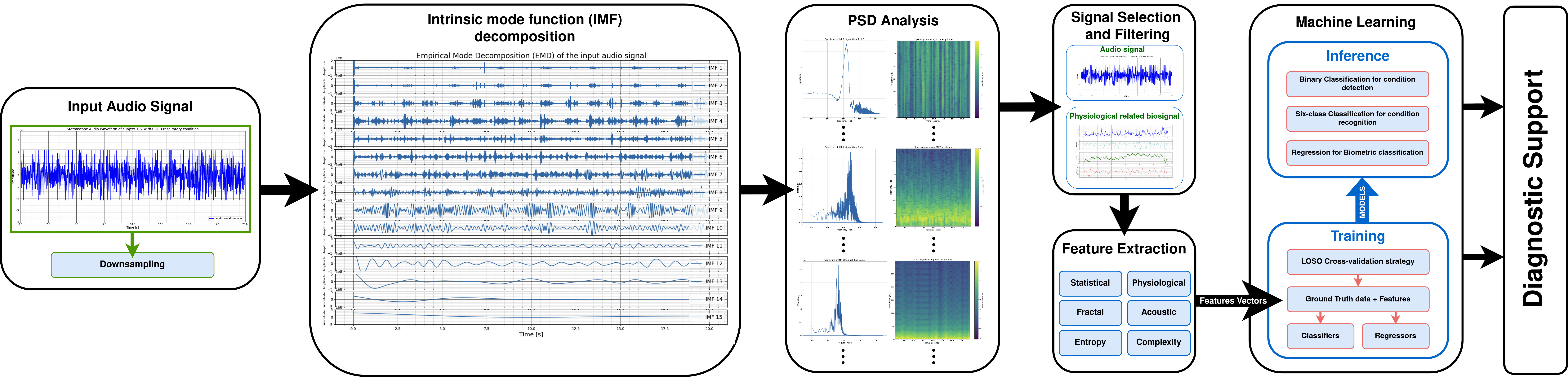}
 \end{center}
\vspace{-3mm}
 \caption{Multi-step methodology for automated diagnosis support for respiratory diseases using stethoscope acoustic signals.}
 \label{fig:methodology_pipeline}
\vspace{-3mm}
\end{figure*}

This work presents a novel, multi-step analytical framework for diagnosing respiratory conditions based on stethoscope-captured audio signals. The framework integrates binary and multi-class classification tasks to identify general health status and specific respiratory pathologies, respectively. Additionally, the study uses regression analysis to estimate patient age and Body Mass Index (BMI), and a secondary classification to determine patient sex. The methodology employs audio signals from stethoscopic recordings and physiologically-related signals derived from them. The analytical pipeline for classification and regression tasks begins with the acquisition of audio signals and their downsampling to a standardized frequency. This is followed by the application of Empirical Mode Decomposition (EMD)~\cite{huang1998EMD} to extract intrinsic mode functions (IMFs) from the acoustic signal. Subsequent steps include conducting power spectral density (PSD) analysis for each IMF, selecting and filtering IMFs based on their power in frequency bands correlated with heart rate and respiration, and extracting features from six selected signals, including the original audio signal. The process concludes with model training using these features and evaluating the performance using established metrics, as depicted in Figure \ref{fig:methodology_pipeline}.

\subsection{Preprocesssing}

In our study, preprocessing is a critical step for efficient feature extraction and the subsequent model training, using audio signals recorded by contact-based stethoscopes with different recording frequencies, depending on the specifications of the medical device used. To ensure efficient computation and storage, these signals are downsampled to a standardized frequency. A Hamming-windowed low-pass finite impulse response (FIR) filter is applied before downsampling to reduce potential signal distortion (aliasing) and high-frequency noise.  FIR filters are selected for their linear phase response, which offers advantages in maintaining temporal characteristics, crucial for subsequent steps.  After filtering, EMD technique is applied to decompose audio signals into IMFs, as detailed by Huang et al. \cite{huang1998EMD}. EMD is particularly effective for the analysis of non-linear and non-stationary signals, characteristic of physiological data, by adaptively decomposing complex signals into simpler components. This method enhances the analysis of physiological phenomena by isolating inherent oscillatory modes within the signal. Subsequently, power spectral density (PSD) analysis is performed on each IMF. This analysis quantifies the distribution of power across frequency bands, enabling precise identification of frequencies associated with physiological activities such as heart rate and respiration. Notably, IMFs are selected for their notable energy within frequency ranges essential for monitoring physiological events: heart rate (1 to 5 Hz), respiration (0.1 to 1.5 Hz), and additional bands (5 to 10 Hz, 10 to 20 Hz, 20 to 100 Hz, and 250 to 1000 Hz) for a wider physiological analysis. Subsequently, two envelope signals are constructed from the final IMFs. The first envelope signal, focusing on lower frequencies, targets respiration information, while the second, derived from higher frequencies, is associated with cardiac activity. This method results in the extraction of six unique signals for feature extraction, where five signals capture physiological characteristics, and the sixth retains its original acoustic nature.

\subsection{Feature extraction}

Following preprocessing, biosignals are systematically segmented into overlapping 10-second windows with a 1-second shift, resulting in a 90\% overlap approximately. This overlap is intentionally designed to prevent information loss during the segmentation phase. Following this, a comprehensive feature extraction process is implemented, which includes not only conventional statistical features in the temporal and frequency domains but also extends to a wider range of informative attributes, following a similar approach than in \cite{Alvarez2023DepressionRPPGs}. The extracted features include vital signs, physiological attributes, morphological characteristics, entropy-based indicators, such permutation and spectral entropy, and fractal dimensions. Additionally, MFCC features are used to capture the acoustical nuances of the original audio signal. The feature set comprises 550 attributes extracted from the six signals associated with each patient. These features also include basic statistical parameters such as mean, minimum, maximum, and standard deviation, alongside dynamic range and signal-to-noise ratio. Time-domain and frequency-domain metrics like slope, total spectral power, and energy are also incorporated. Fractal analysis is represented by Katz, Higuchi, and Petrosian fractal dimensions, among others. Complexity features capture aspects like complex tolerance and delay. Related to heart and respiration indicators, our feature extraction focuses on up to 70 physiological attributes, largely centered around Heart Rate Variability (HRV) metrics across multiple domains. Well-known Python libraries such as Numpy, Antropy \cite{AntropyPythonPackage}, Neurokit2, Librosa and HeartPy \cite{HeartPyVANGENT2019} are utilized to compute all these features. We also introduce custom features specifically tailored to breathing patterns, leveraging both time and frequency domains. Ultimately, the feature extraction process yields a multidimensional and highly discriminative feature set that is instrumental for accurate physiological condition identification within the scope of our proposed methodology.

\subsection{Classification and Regression Models}

In this study, we apply the Multi-Layer Perceptron (MLP) classifier for binary and multiclass classification tasks to differentiate between patients with respiratory conditions and healthy individuals. The MLP is chosen for its ability to handle complex, non-linear relationships in high-dimensional data, which aligns with the extensive feature set extracted from the biosignals. For regression tasks, where we aim to predict continuous variables such as age and BMI, we choose eXtreme Gradient Boosting (XGBoost) regression. We select this method for its superior performance over traditional methods, benefiting from its advanced ensemble learning technique. To ensure consistency and the possibility of replication, we use the default configurations of these algorithms from the \textit{Scikit-learn} library. Before training the models, we conduct a thorough preprocessing phase to handle potential data quality issues. This involves filling in missing values using preceding valid values or the column mean, applying noise reduction and outlier elimination techniques, and normalizing or standardizing the data to ensure consistency across features. These preparatory steps are essential for optimizing the performance of our selected models, leveraging the strengths of MLP for classification and XGBoost for regression.

\section{Experimental setup}

The empirical foundation of this study relies on the ICBHI17 respiratory sound database \cite{rocha2018ICBHIDatabase}, which has been widely utilized in respiratory disease research since its introduction at the 2017 International Conference on Biomedical and Health Informatics (ICBHI). Collected from multiple locations, including Portugal and Greece, the dataset provides a broad representation of subjects.

\subsection{Benchmark Database}

The ICBHI17 database comprises 5.5 hours of audio recordings from 126 subjects (79 males, 46 females, 1 other), resulting in 920 annotated audio files. The database has an average age of 42.99 years, with a standard deviation of 32.21 years. Additionally, it includes 49 subjects who are under the age of 16. These recordings include 6898 respiratory cycles, with 1864 cycles featuring crackles, 886 with wheezes, and 506 exhibiting both. The dataset covers eight respiratory conditions, such as COPD, Lower or Upper Respiratory Tract Infection (LRTI, URTI), or bronchiectasis, as shown in Table \ref{tab:icbhi_database_summary}. Note that asthma was not included in our experiments due to limited samples. An exceptional feature of the ICBHI17 dataset is its robustness, stemming from the use of multiple medical-grade recording devices, mimicking real clinical settings and adding complexity to classification tasks. For consistency, all recordings were resampled to 4 kHz, aligning various original sample rates (44.1 kHz, 10 kHz, and 4 kHz) and ensuring both resolution and compatibility.


\begin{table}[ht!]
\caption{ICBHI Database Summary: Respiratory conditions, healthy subjects, and class labels for 6-class classification \cite{rocha2018ICBHIDatabase}.}
\vspace{-3mm}
\setlength{\tabcolsep}{1.1em}
\def\arraystretch{1.1}
\begin{center}
\begin{tabular}{|l|c|c|c|c|}
\hline
\textbf{Resp. Condition} & \textbf{Patients} & \textbf{Male} & \textbf{Female} & \textbf{Label} \\
\hline
Healthy        & 26 & 13 & 13 & 0 \\
Pneumonia      & 6 & 4 & 2 & 1 \\
Bronchiolitis  & 6 & 4 & 2 & 2 \\
Bronchiectasis & 7 & 2 & 5 & 3 \\
COPD           & 64 & 48 & 15 & 4 \\
URTI           & 14 & 6 & 8 & 5 \\
LRTI           & 2 & 2 & 0 & 6 \\
Asthma         & 1 & 0 & 1 & 7 \\
\hline
Total   & 126 & 79 & 46 & -\\
\hline
\end{tabular}
\end{center}

\label{tab:icbhi_database_summary}
\vspace{-5mm}
\end{table}


\subsection{Protocol and Metrics}

We evaluated our models using a Leave-One-Subject-Out (LOSO) cross-validation approach. For each classification task, we trained either 100 or 126 unique classifiers. In each case, one subject was intentionally excluded from the training set for testing. To evaluate model performance, we averaged results across three metrics: Accuracy, Balanced Accuracy, and F1-score. Accuracy reflects the overall correct predictions but can be misleading in imbalanced datasets like this. Balanced Accuracy addresses this by considering each performance of each class. F1 score emphasizes the balance between precision and recall, making it suitable when the positive class is crucial. Additionally, we also use confusion matrices to assess classification performance. These matrices provide insights into the models' sensitivity, specificity, and overall accuracy for each classification task, highlighting their strengths and limitations in various diagnostic scenarios. For regression tasks, we also used LOSO cross-validation, and we measured the performance of the models using Mean Absolute Error (MAE), Root Mean Squared Error (RMSE), and the coefficient of determination (R\textsuperscript{2}), which is considered more informative and accurate than other metrics in real medical scenarios \cite{chicco2021coefficient}.



\section{EXPERIMENTS AND ANALYSIS}
In this section, we evaluate the performance of our proposed modality and approach through a series of experiments in the benchmark database~\cite{rocha2018ICBHIDatabase}.

\subsection{Classification for Respiratory Conditions}

To evaluate the performance of the proposed methodology, we perform several classification tasks for different respiratory conditions scenarios, as shown in Table \ref{tab:classification_results}.

\begin{table}[ht!]
\vspace{-1mm}
\caption{Summary of the classification models performance.}
\vspace{-3mm}
\setlength{\tabcolsep}{0.7em}
\def\arraystretch{1.1}
\begin{center}
\begin{tabular}{|l|c|c|c|c|}
\hline
\textbf{Classification task} & \textbf{Features type} & \textbf{ACC} & \textbf{BA} & \textbf{F1S} \\
\hline
Binary-Base (PvsH)  &  Au  & 82.52\% & 68.31\% & 89.09\% \\
Binary-Base (COPD)      &  Au   & 86.45\% & 86.45\% & 86.71\% \\

4-classes (RespCond)         & Au  & 77.44\% & 58.49\% & 76.24\% \\
6-classes  (RespCond)        & Au & 72.13\% & 42.45\% & 71.29\% \\
Binary-Base  (Gender)      & Au & 64.20\% & 56.28\% & 73.48\% \\
\hline
Binary (PvsH)     & Au + Phys & 89.68\% & 89.39\%  & 93.19\% \\
Binary (COPD)        &  Au + Phys & 94.43\% & 94.43\% & 94.31\% \\
4-classes (RespCond)         & Au + Phys  & 84.83\% & 71.67\% & 85.11\% \\
6-classes  (RespCond)        & Au + Phys & 75.73\% & 54.81\% & 73.39\% \\
Binary  (Gender)              &   Au + Phys & 75.67\% & 71.29\% & 79.29\% \\

\hline
\end{tabular}
\end{center}

\label{tab:classification_results}
\footnotesize{Abbreviations used in the table - \textit{ACC}: accuracy; \textit{BA}: balanced accuracy; \textit{Au}: 50 Audio features (stats + MFCCs); \textit{Phys}: Physiological related features; \textit{F1S}: F1 score; \textit{PvsH}: pathological versus healthy task; \textit{RespCond}: respiratory conditions recognition task.}
\end{table}

\begin{figure}[ht!]
\vspace{-1mm}
 \begin{center}
   \includegraphics*[width=0.49\textwidth]{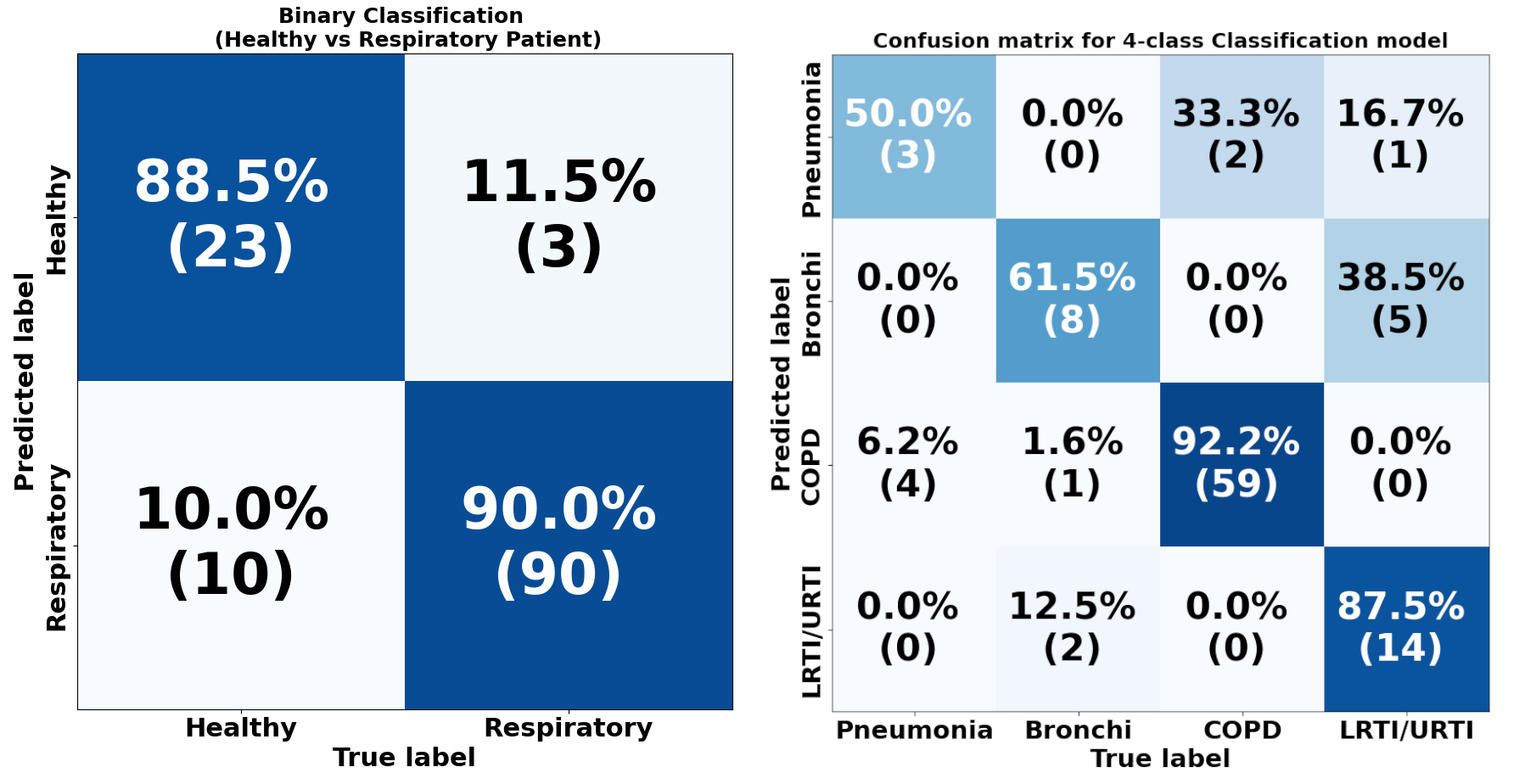}
 \end{center}
 \vspace{-3mm}
 \caption{Confusion matrices illustrating binary classification (left) and 4-class respiratory infection recognition (right), demonstrating model accuracy and differentiation among health statuses and specific infections. The matrices display normalized values in percentages and the distribution of subjects across each class, highlighting dataset imbalances.}
 \label{fig:binary_roc}

\end{figure}

We trained two binary classification models: the first differentiates healthy individuals from those with respiratory conditions, while the second identifies COPD, a chronic lung disease influenced by allergies, smoking, and air pollution, from conditions like pneumonia or bronchiolitis. These models achieved balanced accuracies of 89.39\% and 94.43\%, respectively. For comparison, we trained baseline models using solely audio-based features (statistical and MFCC), which resulted in lower balanced accuracies of 68.31\% and 86.45\%, respectively. Additionally, the binary models exhibited strong discriminative performance, as evidenced by the confusion matrix in Figure \ref{fig:binary_roc}, yielding results on par with state-of-the-art deep learning approaches \cite{wall2022deep}.

We also trained two advanced models to recognize various respiratory conditions. The first, a six-class classifier, focuses on identifying different respiratory conditions from the ICBHI database, excluding asthma and healthy subjects. This model achieved a modest balanced accuracy of 54.81\%, when evaluating conditions based on the average window predictions for each individual, but showing improvement over models using only audio-based features, especially when predictions for each individual were averaged over time windows. For the four-class classification task, we implemented a grouping strategy to address class imbalance and overlapping symptoms by combining URTI with LRTI, and bronchiolitis with bronchiectasis into single categories. Despite these efforts, the performance increase was modest due to the dataset imbalance (COPD cases represent 64\% of conditions), resulting in a balanced accuracy of 71.67\%. This accuracy still marks a noticeable improvement compared to models trained solely on audio-based features, as shown in its confusion matrix, depicted in Figure \ref{fig:binary_roc}.


\subsection{Gender recognition}
Understanding sex differences in medical diagnostics can provide customized treatment and better health outcomes. We have trained an MLP binary classifier using the 550 physiological features derived from acoustic stethoscope data aiming to perform sex recognition. In this experiment, subject 223 was excluded due to the absence of gender label. The model achieved a balanced accuracy of 71\% and a F1-score of 79\%, as depicted in Table \ref{tab:classification_results}. This promising outcome suggests that leveraging such physiological features could further refine classification methods for respiratory conditions through the creation of stratified models, or by incorporating gender as a key variable input. The results underscore the feasibility of audio-based recognition of biometric traits to enhance the potential of personalized medical applications.

\subsection{Regression of Age and BMI}

Additionally, an XGBoost regressor was used for BMI and age estimation from acoustic stethoscope data, as detailed in Table \ref{tab:predictedbmi}. The BMI prediction model achieved a MAE of 3.84 and an R\textsuperscript{2} value of 94.77\%. The age prediction model showed a MAE of 4.66 and an R\textsuperscript{2} of 92.20\%, influenced by the limited representation of individuals aged 20 to 45 years in the dataset. Considering the age variation within the dataset, with a standard deviation of 32.21 years, the performance of the age prediction model, having an MAE standard deviation of 3.97, appears reasonable. Predicting BMI and age from acoustic data signifies notable advancements in medical diagnostics, facilitating preventive strategies for conditions such as obesity and malnutrition and improving patient risk assessment based on age.

\begin{table}[ht!]
\vspace{-3mm}
\caption{Biometric estimation performance of BMI and Age}
\setlength{\tabcolsep}{2.2em}
\def\arraystretch{1.1}
\begin{center}

\begin{tabular}{c|c|c|c|}
\cline{2-4}

& \multicolumn{1}{c|}{\textbf{MAE ± SD}} & \multicolumn{1}{c|}{\textbf{R\textsuperscript{2}}} & \multicolumn{1}{c|}{\textbf{RMSE}} 
\\ \hline

\multicolumn{1}{|l|}{BMI}    & 3.84 ± 3.14   & 0.9475 & 5.41 \\ 
\hline
\multicolumn{1}{|l|}{Age}    & 4.66 ± 3.97  & 0.9220 & 6.12 \\

\hline
\end{tabular}
\end{center}

\label{tab:predictedbmi}

\end{table}


\section{Conclusion}

Our study introduced a framework to train classification models using audio signals obtained from auscultation techniques. Utilizing the well-known ICBHI17 respiratory sound database~\cite{rocha2018ICBHIDatabase}, we extracted clinically relevant biosignals from acoustic data using EMD and spectral analysis techniques.  A unique aspect of our study is the extraction of physiological-related features from these biosignals, offering additional insights into health status beyond traditional audio-based features. Our binary model achieves balanced accuracies of 89\% for patient identification and 72\% for categorizing specific conditions such as Pneumonia and COPD. Despite the heavy imbalance of the dataset, the model still demonstrates robust differentiation capabilities, reflected by the confusion matrices in binary and multi-class tasks. Uniquely, we show that using the same acoustic information we can also estimate age, sex, and BMI. Automatically determining this biometric traits could potentially pave the way for stratified, specific, more accurate models.  We believe that our method constitutes a significant step toward developing unobtrusive, portable, and precise diagnostic tools, with immediate implications for managing respiratory diseases, although future work should focus on proper clinical validation in more varied settings.

\section*{Acknowledgment}

This research is supported by the Academy of Finland 6G Flagship program (Grant 346208) and PROFI5 HiDyn (Grant 326291). The authors wish to acknowledge CSC-IT Center for Science, Finland, for computational resources.

\bibliographystyle{IEEEtran}
\bibliography{references}

\begin{thebibliography}{10}

\bibitem{momtazmanesh2023globalRespiratory}
Sara Momtazmanesh, Sahar~Saeedi Moghaddam, Seyyed-Hadi Ghamari, Elaheh~Malakan
  Rad, Negar Rezaei, Parnian Shobeiri, Amirali Aali, Mohsen Abbasi-Kangevari,
  Zeinab Abbasi-Kangevari, Michael Abdelmasseh, et~al.,
\newblock ``Global burden of chronic respiratory diseases and risk factors,
  1990--2019: an update from the global burden of disease study 2019,''
\newblock {\em EClinicalMedicine}, vol. 59, 2023.

\bibitem{kim2022comingStethoscope}
Yoonjoo Kim, YunKyong Hyon, Sunju Lee, Seong-Dae Woo, Taeyoung Ha, and Chaeuk
  Chung,
\newblock ``The coming era of a new auscultation system for analyzing
  respiratory sounds,''
\newblock {\em BMC Pulmonary Medicine}, vol. 22, no. 1, pp. 119, 2022.

\bibitem{rocha2018ICBHIDatabase}
BM~Rocha, Dimitris Filos, L~Mendes, Ioannis Vogiatzis, Eleni Perantoni,
  Evangelos Kaimakamis, P~Natsiavas, Ana Oliveira, C~J{\'a}come, A~Marques,
  et~al.,
\newblock ``A respiratory sound database for the development of automated
  classification,''
\newblock in {\em Precision Medicine Powered by pHealth and Connected Health:
  ICBHI 2017, Thessaloniki, Greece, 18-21 November 2017}. Springer, 2018, pp.
  33--37.

\bibitem{Reichert2008DSPRespiratory}
Sandra Reichert, Raymond Gass, Christian Brandt, and Emmanuel Andrès,
\newblock ``Analysis of respiratory sounds: State of the art,''
\newblock {\em Clinical Medicine : Circulatory, Respiratory and Pulmonary
  Medicine}, vol. 2, 05 2008.

\bibitem{Fattahi2022Filtering}
Davood Fattahi, Reza Sameni, Ethan Grooby, Kenneth Tan, Lindsay Zhou, Arrabella
  King, Ashwin Ramanathan, Atul Malhotra, and Faezeh Marzbanrad,
\newblock ``A blind filtering framework for noisy neonatal chest sounds,''
\newblock {\em IEEE Access}, vol. 10, 04 2022.

\bibitem{lin2015automaticDSPResp}
Bor-Shing Lin, Huey-Dong Wu, Sao-Jie Chen, et~al.,
\newblock ``Automatic wheezing detection based on signal processing of
  spectrogram and back-propagation neural network,''
\newblock {\em Journal of healthcare engineering}, vol. 6, pp. 649--672, 2015.

\bibitem{bahoura2009patternGMMs}
Mohammed Bahoura,
\newblock ``Pattern recognition methods applied to respiratory sounds
  classification into normal and wheeze classes,''
\newblock {\em Computers in biology and medicine}, vol. 39, no. 9, pp.
  824--843, 2009.

\bibitem{aykanat2017classificationSVM}
Murat Aykanat, {\"O}zkan K{\i}l{\i}{\c{c}}, Bahar Kurt, and Sevgi Saryal,
\newblock ``Classification of lung sounds using convolutional neural
  networks,''
\newblock {\em EURASIP Journal on Image and Video Processing}, vol. 2017, no.
  1, pp. 1--9, 2017.

\bibitem{kok2019novelDWTMFCC}
Xuen~Hoong Kok, Syed~Anas Imtiaz, and Esther Rodriguez-Villegas,
\newblock ``A novel method for automatic identification of respiratory disease
  from acoustic recordings,''
\newblock in {\em 2019 41st Annual International Conference of the IEEE
  Engineering in Medicine and Biology Society (EMBC)}. IEEE, 2019, pp.
  2589--2592.

\bibitem{Sibghatullah2021COPDdetection}
Sibghatullah~I. Khan and Ram~Bilas Pachori,
\newblock ``Automated classification of lung sound signals based on empirical
  mode decomposition,''
\newblock {\em Expert Systems with Applications}, vol. 184, pp. 115456, 2021.

\bibitem{mukherjee2021automaticLPCC}
Himadri Mukherjee, Priyanka Sreerama, Ankita Dhar, Sk~Md Obaidullah, Kaushik
  Roy, Mufti Mahmud, and KC~Santosh,
\newblock ``Automatic lung health screening using respiratory sounds,''
\newblock {\em Journal of Medical Systems}, vol. 45, pp. 1--9, 2021.

\bibitem{BARDOU2018CNNsResp}
Dalal Bardou, Kun Zhang, and Sayed~Mohammad Ahmad,
\newblock ``Lung sounds classification using convolutional neural networks,''
\newblock {\em Artificial Intelligence in Medicine}, vol. 88, pp. 58--69, 2018.

\bibitem{wall2022deep}
Conor Wall, Li~Zhang, Yonghong Yu, Akshi Kumar, and Rong Gao,
\newblock ``A deep ensemble neural network with attention mechanisms for lung
  abnormality classification using audio inputs,''
\newblock {\em Sensors}, vol. 22, no. 15, pp. 5566, 2022.

\bibitem{huang1998EMD}
Norden~E Huang, Zheng Shen, Steven~R Long, Manli~C Wu, Hsing~H Shih, Quanan
  Zheng, Nai-Chyuan Yen, Chi~Chao Tung, and Henry~H Liu,
\newblock ``The empirical mode decomposition and the hilbert spectrum for
  nonlinear and non-stationary time series analysis,''
\newblock {\em Proceedings of the Royal Society of London. Series A:
  mathematical, physical and engineering sciences}, vol. 454, no. 1971, pp.
  903--995, 1998.

\bibitem{Alvarez2023DepressionRPPGs}
Constantino Álvarez Casado, Manuel Lage~Cañellas, and Miguel Bordallo~López,
\newblock ``Depression recognition using remote photoplethysmography from
  facial videos,''
\newblock {\em IEEE Transactions on Affective Computing}, pp. 1--13, 2023.

\bibitem{AntropyPythonPackage}
Raphael Vallat and Matthew Walker,
\newblock ``An open-source, high-performance tool for automated sleep
  staging,''
\newblock {\em eLife}, vol. 10, 10 2021.

\bibitem{HeartPyVANGENT2019}
Paul {van Gent}, Haneen Farah, Nicole {van Nes}, and Bart {van Arem},
\newblock ``Heartpy: A novel heart rate algorithm for the analysis of noisy
  signals,''
\newblock {\em Transportation Research Part F: Traffic Psychology and
  Behaviour}, vol. 66, pp. 368--378, 2019.

\bibitem{chicco2021coefficient}
Davide Chicco, Matthijs~J Warrens, and Giuseppe Jurman,
\newblock ``The coefficient of determination r-squared is more informative than
  smape, mae, mape, mse and rmse in regression analysis evaluation,''
\newblock {\em PeerJ Computer Science}, vol. 7, pp. e623, 2021.

\end{thebibliography}

\end{document}